\documentclass[aps,pre,twocolumn,epsfig,floatfix,twoside,showpacs]{revtex4}
\usepackage{epsfig}

\begin{document}
\input epsf
\title{Spinodal of supercooled polarizable water}
\author{
P.~Gallo$^\dagger$\footnote[1]{Author to whom correspondence  
should be addressed; e-mail: gallop@fis.uniroma3.it}, 
M. Minozzi$^{\dagger \ddagger}$ and
M.~Rovere$^\dagger$ }
\affiliation{$\dagger$ Dipartimento di Fisica, 
Universit\`a ``Roma Tre'', \\ 
and Democritos National Simulation Center,\\ 
Via della Vasca Navale 84, 00146 Roma, Italy \\
$\ddagger$ CNR Istituto di Acustica ``O. M. Corbino'', Via 
del Fosso del Cavaliere 100, 00133 Roma, Italy }

\begin{abstract}
We develop a series of molecular dynamics computer simulations 
of liquid water, performed with a polarizable potential model, 
to calculate the spinodal line and the curve of 
maximum density inside 
the metastable supercooled region.
After analysing the structural properties,
the liquid spinodal line is followed down to $T=210$~K.
A monotonic decrease is found in the explored region.
The curve of maximum density bends on approaching the spinodal line.
These results, in agreement with similar studies on non polarizable
models of water, are consistent with
the existence of a second critical point for water. 
\end{abstract}

\maketitle

\section{introduction}

Liquid water behaviour as a function of temperature and pressure
differs from that of most liquids.
As a major example the presence of a line of maximum 
density (TMD line) in the P-T phase diagram
is most significant,
as it divides the entire P-T phase diagram into two regions with remarkably 
different properties: the coefficient of thermal expansion is negative on the
low-temperature side of the TMD line, while it is positive on the 
high-temperature side. 
Similarly, several other thermodynamic and dynamical quantities show 
peculiar properties.
For instance the isothermal compressibility and heat 
capacity exhibit a minimum as a function of temperature and
show an anomalous behavior more pronounced in the
supercooled liquid state~\cite{angell_review}. 
When decreasing temperature in fact the 
coefficient of thermal expansion, the isothermal compressibility 
and the constant-pressure specific 
heat increase rapidly.
These quantities appear to diverge at a temperature 
$T\approx -45~C $ 
if they are  extrapolated below the lowest 
temperatures at which they are measurable 
($T=-42$~C at $P=1$~$bar$)~\cite{angell_review}.
The supercooled region below $T=-42$~C is experimentally
unreachable due to the strong tendency of water to crystallize.
It is nonetheless believed that nucleation in supercooled water might be due
to the presence of impurities that drive the liquid toward the more stable
phase~\cite{debenedetti}. Therefore the experimentally unreachable 
zone of supercooling is believed to be physically significant.
Measurements of the rate of evaporation on amorphous water have in fact proved 
that the amorph can be connected with normal liquid water 
by a reversible thermodynamic path at 
atmospheric pressure~\cite{speedydebenedetti}.
Besides at ambient pressure experiments have proven the existence
of supercooled liquid water close to the glass transition 
temperature~\cite{kay}.

In spite of all the interest driven by these anomalies a
coherent theory of the thermodynamic and transport properties of supercooled 
water does not yet exist, also due to the difficulties encountered
in experiments. 

Different thermodynamic scenarios have been proposed 
through the years for the peculiar 
metastable behaviour of water~\cite{pabloreview}. Among these three 
of them have received a great attention in literature.
The stability limit conjecture (SLC) is the first scenario proposed.
It attributes the anomalies of water to the presence
of a continuous retracing spinodal curve, bounding the superheated and 
supercooled states. The spinodal, which represents
the limit of mechanical stability of a liquid, is hypothized 
to retrace to higher pressure 
values below a temperature at which it intersects the locus of the
TMD~\cite{angell-speedy,Speedy}.
The second critical point scenario (SCP), 
based on extrapolation of simulated data,
ascribes the anomalous properties of water to the presence of a metastable, low
temperature liquid-liquid critical point, associated with a phase transition
between a low-density and a high density liquid phase~\cite{nature1}. 
Experimental evidences of the existence of two liquid phases
compatible with a second critical point have
been presented~\cite{mishima}.
The singularity free scenario (SF)
explains the behaviour of supercooled water with the presence of
anomalous fluctuations due to the hydrogen bonds 
and no underlying singularity is invoked~\cite{Sastry}.
The SF assumes
that upon isobaric cooling the thermodynamic response functions go
through a maximum but remain finite. An anomalous increase in
isothermal  compressibility, heat capacity, and thermal expansion is
explained by the existence of the 
density maxima locus, which is negatively sloped in the 
$P,T$ plane~\cite{debenedetti}.
According to this scenario, no phase transition or critical point
occurs at low temperatures.

A fundamental role in the clarification of the
scenario of supercooled water is played by the 
study of the behaviour of the spinodal line,
and its relation with the line of TMD.
Due to the limitations of the experiments in the supercooled realm,
in order to assess the different hypothesis
numerical studies have become of uttermost importance.
Many computer simulation studies indeed have been 
performed with several different water site models, 
ST2,TIP4P,SPC/E and TIP5P~\cite{spinodal1,Essmann,poole,tanaka,hpss,Mossa}. 
All the above models do not take into account explicitly the polarizability 
of the water molecules. 
Polarizable models for water have been also 
developed~\cite{pol1,pol2,pol3,pol4,pol5,r+s1,bsv1,pol6,pol7,pol8,pol9,pol10,pol11} to give a more realistic 
description of the behaviour of the system
in different thermodynamic conditions.
Depending on the zone of the phase diagram investigated they have
proven to be equally or more realistic to describe
the features of water. Studies of the behaviour of the
spinodal line for polarizable water model potential
have been never carried out.
It is therefore of interest to study the behaviour of
the spinodal line for potentials where polarizability
is explicitly taken into account.

We present here a Molecular Dynamics, MD, study of 
the spinodal and the TMD line of supercooled water 
performed with the 
polarizable potential, introduced in the literature
by Ruocco and Sampoli~\cite{r+s1} and parametrized by
Brodholt, Sampoli and Vallauri~\cite{bsv1}, the BSV potential.
We found that the BSV model is appropriate for 
our study since
it reproduces the thermodynamical properties 
of water in the region of the gas-liquid coexistence
better than  other polarizable models 
over a broad temperature range~\cite{vallauri05}.  
It has also been shown that within the BSV model 
the site-site radial distribution functions are
in good agreement with the experimental data  
in a broader range of thermodynamical 
conditions.~\cite{jedlo1,vall1,conf1,tmd1,tmd2}

In the next section we describe the computer
simulation in details. After presenting in the third section the structural
properties of the system upon supercooling,
the behaviour of the spinodal and of the TMD lines 
are reported in the fourth section. Last section is devoted to conclusions.

\section{computer simulations}

We performed MD computer simulations of water 
with the Model 4 of ref.~\cite{bsv1}.
An induced polarizable dipole moment $P$, located in the center-of-mass of the 
molecule, describes the effect of the electric field of the environment on the 
molecules. The induced dipole $p_i=\alpha E_i$
is calculated from the local electric field $E_i$ with
an iterative procedure
by assuming an isotropic polarizability. The value of
this polarizability is fixed to the single molecule value
$\alpha=1.44$~\AA$^3$~\cite{r+s1}.

Each simulation is conducted in the $NVT$ ensemble
with  $256$ water molecules enclosed in a cubic box with 
periodic boundary conditions. 
The simulations have been performed with the
minimum image convention and a cut-off of the interactions
at half of the box length. 
The long range part of the
electrostatic interactions is taken into account 
with the reaction field method. 
Details of the extension of the reaction field method
to include polarization effects can be found in references~\cite{r+s1,r+s2}. 
The time step $\delta t$ for the integration of the molecular trajectories is 
fixed at $1$~$fs$.

To analyse the phase diagram and the spinodal curve inside the metastable 
supercooled region we carried out isothermal simulations for eight different
temperatures: 
$350$, $300$, $280$, $260$, $240$, $230$, $220$ and $210$~K. For each  
eight different densities are simulated, namely
$1.05$, $1.00$, $0.98$, $0.95$, $0.90$, $0.87$, $0.85$ and
$0.83$~$gr/cm^3$.
The box length $L$ spans from $L= 19.39$~\AA\ to $L=20.96$~\AA\ to
cover the range of investigated densities. 

During equilibration 
the temperature is controlled via a velocity rescaling procedure. 
Production runs are performed in the microcanonical ensemble.
For the temperatures and the densities closer to the spinodal line
the longest equilibration time required was of $5$~$ns$.

In  tables \ref{table1} and \ref{table2} we report the temperatures,
densities, pressures and internal energies of the 
simulations we have performed.

\section{Structural properties}

\begin{figure}[h]
\centerline{\psfig{file=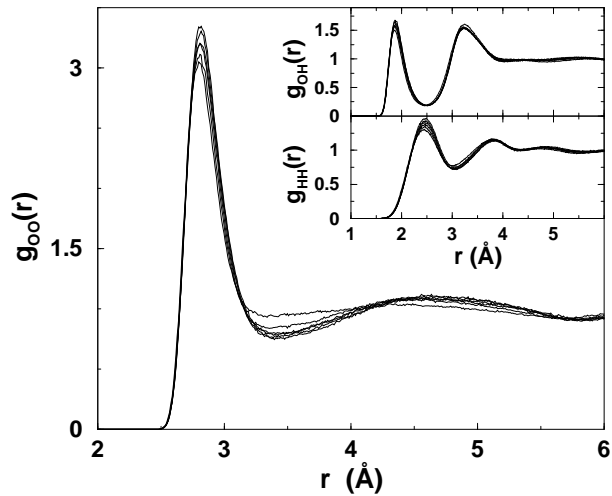,width=8.0cm,clip=!}}
\caption{Site-site 
pair correlation functions of the BSV water at $T=300$~K and densities 
$\rho=1.00$, $0.98$, $0.95$, $0.90$, $0.87$, $0.85$, $0.83$.
$g_{OO}(r)$ is in the main frame. The insets show
$g_{OH}(r)$ and $g_{HH}(r)$ starting from the top.
}
\label{fig:1}
\end{figure}

\begin{figure}[h]
\centerline{\psfig{file=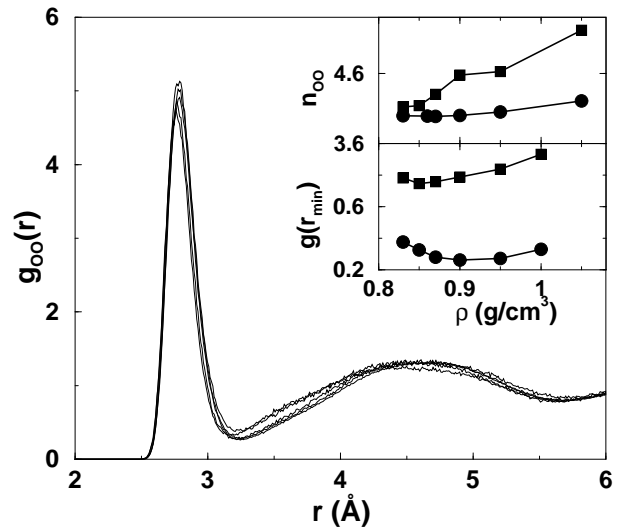,width=8.0cm,clip=!}}
\caption{
Pair correlation function
$g_{OO}(r)$ at $T=220$~K and the same densities as Fig.~\ref{fig:1}. 
In the inset on the top
number of nearest neighbours $n_{OO}$ as a function of $\rho$ 
for $T=300$~K (full squares)
and $T=220$~K (full circles); in the inset below
value of the first minimum of $g_{OO}(r)$, $g(r_{min})$,
as a function of $\rho$ at $T=300$~K (full squares)
and $T=220$~K (full circles).}
\label{fig:1bis}
\end{figure}

Structural 
quantities have been calculated averaging from a minimum of
$2.5 \times 10^{3}$ configurations at $T=300$~K to a maximum of
$3 \times 10^{3}$ configurations at $T=220$~K, all equally spaced
and taken every $2ps$.
We have calculated the site-site radial distribution functions (RDF)
$g_{\alpha\beta}(r)$ to monitor the 
internal structure at every temperature. We report 
$g_{OO}(r)$, $g_{OH}(r)$ and $g_{HH}(r)$ for $T=300$~K in Fig.\ref{fig:1} and 
$g_{OO}(r)$ for $T=220$~K in Fig.\ref{fig:1bis} calculated at several 
densities. At $T=300$~K the peak positions agree with the experimentally 
measured structure of water~\cite{soper94}. 
Upon decreasing temperature the peaks remain at the same position
and become sharper. 
By varying the densities we do not observe  
very large changes in the $g_{\alpha\beta}(r)$ apart for an increase
in the height of the peaks,
even when the system approaches the spinodal line (see next section).
This is an important test
to verify the absence of disomogeneities or cavitation 
phenomena close to the limit of mechanical stability.
A further check has been performed with the calculations
of the local density fluctuations.
For this sake we have
divided our simulation box into $64$ 
sub-boxes and calculated the average distribution functions 
of the local density (not shown). In all cases
we do not find deviations from a gaussian distribution
centered on the fixed density of the system.

A deeper insight in the local ordering of the atoms
can be achieved by calculating 
the coordination numbers.
In the upper inset of Fig.~\ref{fig:1bis} we show $n_{OO}$,   
the coordination number of the oxygens, as a 
function of $\rho$:
\begin {equation}
n_{OO}=4 \pi \rho \int_{0}^{r_{min}}{g_{OO}(r) r^2 dr}
\label{eq:1}
\end{equation}
where $r_{min}$ is the value of the interatomic distance at which the first
minimum in $g_{OO}(r)$ is located. It is important to
observe the different behaviour of $n_{OO}$ at the two temperatures
$T=300$~K and $T=220$~K on lowering the density.   
At $T=300$~K for the highest density 
$n_{OO}$ results to be $5.2$. This value is the signature of
an open network of water, where also interstitial 
molecules are present~\cite{jedlo1}. 
$n_{OO}$ then decreases until it reaches  
the limiting value $4$ characteristic of the local 
tetrahedral order.
At the lower temperature $T=220$~K $n_{OO}$ 
is found to be quite constant around the value $4$.
These findings indicate
that the open network of $H$-bonds becomes better defined as $\rho$ 
and/or $T$ decrease.

In connection with the tendency of the $H$-bond network
to become more ordered it is expected a 
sharpening of peaks of the RDF.
A quantitative measure of this sharpening can be obtained looking at 
the value of the RDF at its first minimum, $g(r_{min})$.
As a consequence we also expect to observe a decrease of $g(r_{min})$
as the coordination shells become better defined. 
In the lower inset of Fig.~\ref{fig:1bis} 
we observe instead the presence of a minimum in the behaviour of
$g(r_{min})$ as function of the density. The minimum is more evident
at $T=220$~K. This effect evidences an important change in the  
thermodynamical properties of the system, since it
is related to the behaviour of the entropy in the vicinity of
the spinodal, as previously discussed in literature~\cite{spinodal1}.
Therefore we can consider that the change of slope in the $g(r_{min})$ 
curve is a signature of the approaching of the system to the
limit of mechanical stability.

\section{Spinodal line}

The limit of the mechanical stability of a system can
be determined from a study of the isothermal compressibility.
In the space of variables identifying the thermodynamic 
state of the system in fact, the condition for the mechanical stability 
is~\cite{debenedetti}
\begin{equation}
K_T>0 \label{eq:kt>0}
\end{equation}
where $K_T$ is the isothermal compressibility defined as
\begin{equation}
K_T=\frac{1}{\rho}\left(\frac{\partial \rho}{\partial P}\right)_T
\label{eq:kt}
\end{equation}
In the phenomenological description of the liquid-gas transition,
as given for instance by mean field theories like the Van der Waals equation,
along an isotherm the limits of the mechanical stability are
marked by the changes in the slope of the $P_T(\rho)$ curve.
Therefore from Eq.~(\ref{eq:kt}) it is found that
the crossing to unstable states with $K_T<0$
takes place at singularity points where $K_T$ diverges.
The spinodal line is identified by these singularity points.
In particular the point where the minimum of $P_T(\rho)$ 
is located represents the limit of stability of the superheated
liquid and so it belongs to the liquid branch of the spinodal. 

We are interested here at the liquid spinodal
which extends in the $P-T$ plane below the liquid-gas coexistence into
the region of negative pressures. The liquid spinodal indicates also 
the limiting values of tension for the existence of an homogeneous
fluid before phenomena like cavitation take place~\cite{debenedetti}.

\begin{figure}[t]
\centerline{\psfig{file=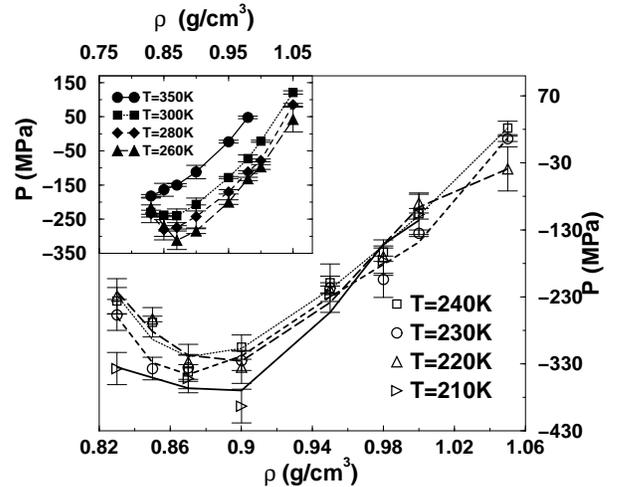,width=8.0cm,clip=!}}
\caption{Isotherms (symbols) for 
densities ranging from 1.05 to 0.83 $gr/cm^3$. 
Errors are calculated
with the method of statistical inefficiency \protect\cite{allen}.
In the inset temperatures range from $350$~K to $260$~K. 
In the main frame temperatures range from $240$~K to $210$~K
and lines are fit with a fourth order polynomial function.}
\label{fig:3}
\end{figure}

The behaviour of $P_T(\rho)$ for the BSV water is shown in
Fig.~\ref{fig:3} for 
$ \rho$ ranging from 1.05 to 0.83 $gr/cm^{3}$. 
In the inset of Fig.~\ref{fig:3}  
four isotherms are represented upon lowering the 
temperature from $T=350$~K to $T=260$~K. 
Isotherms from $T=240$~K to $T=210$~K are 
shown in the main frame of Fig.~\ref{fig:3}. 
From $T=300$~K to the lowest temperature investigated
the $P_T(\rho)$ curves exhibit a minimum in the simulated range
of densities.
The isotherms corresponding
to lower temperatures have larger fluctuations
due to the influence of both the low temperatures and the negative pressures
In order to better follow the curves for the lowest temperatures 
shown in Fig.~\ref{fig:3}
we fitted the calculated points of the isotherms
with a fourth-order polynomial function. We observe however that 
due to the fluctuations
the fit systematically slightly overestimates the values of the minima.
Therefore we will consider the minima directly observed for
the plot of the spinodal line.

Based on Eq.~\ref{eq:kt>0} 
the minima of the isotherms represent the limit
of stability of the superheated liquid and  
identify the values of the pressure at the spinodal, $P_s(T)$.
As $T$ decreases $P_s(T)$ clearly shifts to larger values of $\rho$. 
This behaviour is  similar to that observed for TIP5P~\cite{Mossa} 
while for ST2 and TIP4P this shift is less evident~\cite{spinodal1}. 

\begin{figure}[t]
\centerline{\psfig{file=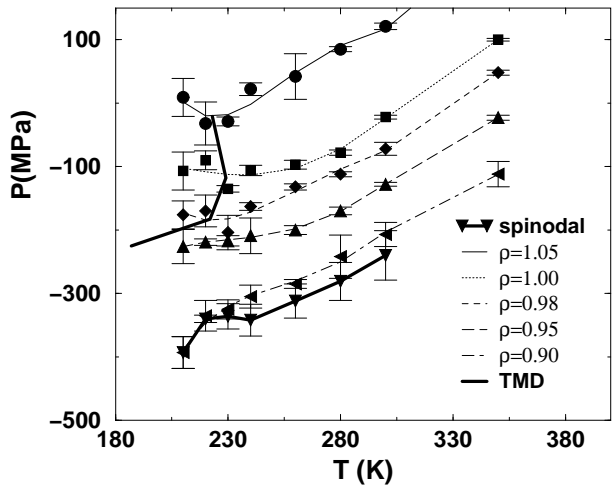,width=8.0cm,clip=!}}
\caption{$P_{\rho}(T)$ isochores for several values of $\rho$ in $g/cm^3$,
spinodal line and TMD curve.
As $\rho$ decreases, approaching the spinodal line, the isochores
exhibit a minimum at smaller values of $T$.
The spinodal line decreases upon lowering temperature and
no retracing behaviour is observed.}
\label{fig:5}
\end{figure}

The resulting estimated liquid
spinodal line $P_s(T)$ is plotted in fig.~\ref{fig:5}. In the 
same figure also the $P_{\rho}(T)$ isochores are shown as obtained by the
data set of tables~\ref{table1} and~\ref{table2}.
At least down to the lowest investigated temperature, $T=210$~K, 
the spinodal line is not retracing. 
In the SLC framework the spinodal line is expected to 
become reentrant at the point where it intersects the TMD line in the 
phase diagram. We recall that this behaviour is required
if the TMD line has a negative slope~\cite{Speedy,debenslc1,spinodal1}.  

Along the TMD line the coefficient of thermal expansion 
\begin{equation}
\alpha_P=\frac{1}{V}\left(\frac{\partial V}{\partial T}\right)_P
\label{eq:alpha}
\end{equation}
goes to $0$.
Since the thermal pressure coefficient $\gamma_V$
\begin{equation}
\gamma_V=\left(\frac{\partial P}{\partial T}\right)_V
\end{equation}
is connected to the coefficient $\alpha_P$ by the following relation
\begin{equation}
\gamma_V=\frac{\alpha_P}{K_T}
\end{equation}
the TMD points are on 
the line connecting the minima of the 
estimated $P_{\rho}(T)$ isochore for different 
$\rho$.

According to Fig.~\ref{fig:5}
as $\rho$ decreases, approaching the spinodal line, 
the isochores exhibit a 
minimum at smaller values of $T$.
 
The isochore for the lowest reported
density $\rho=0.90$~$gr/cm^3$  does not show any minimum in the range
of $T$ investigated 
where instead the minima for the other isochores
are found. 
The TMD line does not approach the
spinodal line and at the lower $T$ investigated 
it results to have a positive slope.
This behaviour of the TMD line prevents an intersection with the spinodal
at negative P and low T, unless a further change in the slope of
the TMD line is hypothized.

These results are in agreement with the findings of 
the studies done on non polarizable models 
for water~\cite{spinodal1,Essmann,poole,tanaka,hpss,Mossa}.
In fact for all these models the TMD line changes slope
to avoid intersection with a non reentrant spinodal.
Differences among the models can be found in the
values of pressures and temperatures where the curves
span. 
In particular the change of slope of the
TMD line appears in our model at 
negative pressures, $P < -120$~MPa,
similar to SPC/E~\cite{hpss}, $P < -80$~MPa and to 
ST2~\cite{spinodal1}.
For TIP5P~\cite{Mossa} the change of slope appears
at ambient pressure instead.

The most studied 
alternative to the SLC for explaining the anomalies
of water is the presence of a second critical point. 
The SCP hypothesis~\cite{nature1,tanaka} is based on the consideration that
in the glassy state two distinct types of structure
are observed~\cite{mishima1}, the low density amorphous (LDA) ice
and the high density amorphous (HDA) ice. This polymorphism
of water would be extended into the liquid phase
and it would emerge as a coexistence line between two liquid phases
culminating in a second critical point~\cite{mishima2}.
Computer simulation 
of waterlike lattice models show evidence in fact of two 
immiscible liquid forms of water~\cite{Sri,Borick,Pablolm1,Pablolm2}.

\begin{figure}
\centerline{\psfig{file=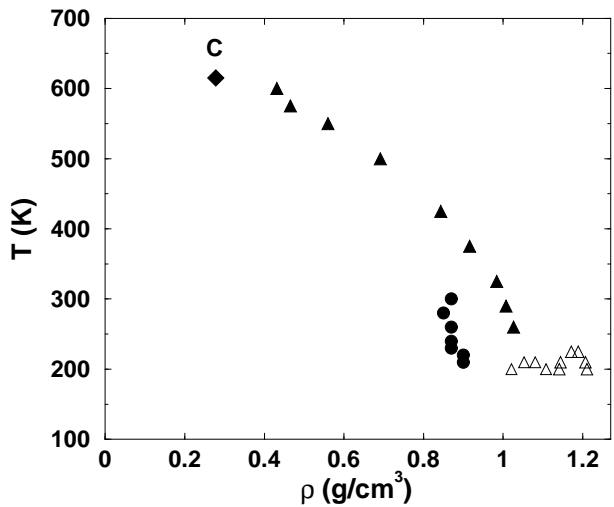,width=8.0cm,clip=!}}
\caption{Liquid branch of the coexistence curve obtained by 
GEMC~\cite{vallauri05} (filled triangles) and the spinodal line of
the present work
(black circles) in the $T,\rho$ plane for the BSV water. The diamond represents
the estimated gas-liquid critical point $C$. Portions of the liquid-liquid 
coexistence regions are shown by the open triangles~\cite{vallauri05}.
}
\label{fig:7}
\end{figure}
More recently 
several amorphous ice phases~\cite{tulk} have been experimentally observed
and signatures have been found in computer simulation of the existence of
several liquid-liquid coexistences in supercooled water~\cite{geiger}.
Also Gibbs ensemble Monte Carlo (GEMC) studies
on the BSV water have found two liquid-liquid coexistence curves
besides the liquid-vapor equilibrium~\cite{vallauri05}.
The liquid branch of the gas-liquid coexistence curve obtained
with the GEMC is reported in Fig.~\ref{fig:7}, where
we collect the relevant features of the
behaviour of the BSV water in the $(T,\rho)$ thermodynamical plane.
The spinodal line obtained in the present work in the supercooled
region is compatible with the GEMC results.
It comes out that the metastability region
explored here goes very close to the
boundaries of the region where new phases of the liquid 
water have been found.  

\section{Conclusions}

We have done Molecular Dynamics simulation of supercooled
water described with a polarizable BSV potential.
The aim of the paper was to calculate the behaviour
of the spinodal line and of the TMD line down to
lowest possible temperatures and pressures.

We have performed a structural analysis of water
described by the BSV polarizable potential
in the region of the supercooled liquid and we have found
that the system persists in an homogeneous phase down to
$T=210$~K, $P \approx -390$~MPa and
this occurs for $\rho \approx 0.9$. Signatures of 
the presence of the mechanical
stability limit are observed in the behaviour of the
values of the RDF at the first minimum.

The boundaries of the liquid mechanical stability at low $T$ and negative $P$
have been determined by the direct calculations of the liquid spinodal
trough the minima of the isotherms $P_T(\rho)$. The line of the TMD
in the same thermodynamical region has also been detected 
from the minima of the isochores $P_\rho(T)$.

We did not find any retracing of the spinodal line in the 
investigated region while we observe a change of slope of the
TMD line which prevents for continuity a possible intersection of
the TMD with the spinodal even down to lower $T$ and $P$. 
These results are in agreement with
previous calculations done with non polarizable site 
potentials~\cite{spinodal1,Essmann,poole,tanaka,hpss,Mossa}.
Our analysis excludes indeed the possibility of the SLC scenario.

As a matter of fact the no retracing of the spinodal is compatible however
with both the SI interpretation~\cite{Sastry} and the possible existence of a
second critical point~\cite{nature1}.
On the other hand we have shown that the liquid spinodal calculated
in the present work is in agreement with the gas-liquid coexistence curve 
obtained with the GEMC~\cite{vallauri05}. 
Since in the GEMC there are clear indications
of the existence of two liquid-liquid coexistence regions we can infer that
our results support the second critical point scenario. 

The use of a polarizable potential in the present work has given
an answer to two important issues. The first is that polarizable and
non polarizable
potentials are equally able to describe an unique, and therefore likely,
scenario for the liquid-gas spinodal of supercooled water. 
This is an important achievement as the parameters of these
potentials have been calculated through fit to experimental data 
performed at ambient conditions. Therefore the uniqueness of
phase behaviour far from these conditions reinforces the
possibility that these potentials are able to predict
existing features of water.
The second is that the analysis
performed with the polarizable model can be extended in a different
zone of the phase diagram for determining the liquid-liquid spinodal
and indications about the second critical point location.

\newpage

\begin{table}
{\small
\setlength{\tabcolsep}{1mm}
\begin{tabular}{|r|r|r|r|}

\hline
\hline
$T~(K)$&$\rho~(g/cm^3)$&$P\pm\delta P~(MPa)$&$U\pm\delta U~(KJ/mol)$\\
\hline
\hline

$350$&$1.00$&$ 100\pm 2$&$-39.16\pm 0.17$\\
$350$&$0.98$&$\ \ 48\pm 24$&$-39.03 \pm 0.18$\\
$350$&$0.95$&$-23\pm 4$&$-37.63\pm 0.04$\\
$350$&$0.90^*$&$-112\pm 40$&$-37.28\pm 0.04$\\
$350$&$0.87$&$-150\pm 4$&$-36.34\pm 0.18$\\
$350$&$0.86^*$&$-159\pm 4$&$-36.65\pm 0.06$\\
$350$&$0.85$&$-164 \pm 38$&$-36.52\pm 0.06$\\
$350$&$0.83$&$-182 \pm 4$ &$-35.58 \pm 0.04$\\

\hline

$300$&$1.05$&$121\pm 5$&$-42.92\pm 0.13$\\
$300$&$1.00$&$-22\pm 3$&$-41.55\pm 0.10$\\
$300$&$0.98$&$-72\pm 10$&$-42.11\pm 0.05$\\
$300$&$0.95$&$-128\pm 20$&$-40.97\pm 0.02$\\
$300$&$0.90^*$&$-207\pm 37$&$-40.76\pm 0.15$\\
$300$&$0.87$&$-240\pm 37$&$-39.72\pm 0.04$\\
$300$&$0.86^*$&$-243\pm 25$&$-40.02\pm 0.07$\\
$300$&$0.85$&$-238\pm 5$&$-39.49\pm 0.04$\\
$300$&$0.83$&$-230\pm 6$&$-39.02\pm 0.16$\\

\hline

$280$&$1.05$&$85\pm 4$&$-44.10\pm 0.14$\\
$280$&$1.00$&$-78\pm 4$&$-43.25\pm 0.04$\\
$280$&$0.98$&$-112\pm 4$&$-43.77\pm 0.08$\\
$280$&$0.95$&$-170\pm 6$&$-42.56\pm 0.06$\\
$280$&$0.90^*$&$-242\pm 34$&$-42.77\pm 0.06$\\
$280$&$0.87$&$-274\pm 5$&$-41.62\pm 0.09$\\
$280$&$0.86^*$&$-280\pm 4$&$-41.71\pm 0.10$\\
$280$&$0.85$&$-281\pm 30$&$-41.06\pm 0.12$\\
$280$&$0.83$&$-234\pm 26$&$-41.57\pm 0.11$\\

\hline

$260$&$1.05$&$42\pm 36$&$-45.90\pm 0.10$\\
$260$&$1.00$&$-97\pm 7$&$-44.91\pm 0.03$\\
$260$&$0.98$&$-132\pm 5$&$-45.67\pm 0.11$\\
$260$&$0.95$&$-200\pm 7$&$-44.47\pm 0.08$\\
$260$&$0.90^*$&$-285\pm 7$&$-44.65\pm 0.10$\\
$260$&$0.87$&$-312\pm 27$&$-44.26\pm 0.08$\\
$260$&$0.86^*$&$-300\pm 21$&$-44.18\pm 0.09$\\
$260$&$0.85$&$-270\pm 30$&$-44.38\pm 0.06$\\
$260$&$0.83$&$-216\pm 28$&$-43.90\pm 0.08$\\
\hline
\hline

\end{tabular}}
\caption{Temperature, density, pressure and internal energy of
the simulated state points.}
\label{table1}
\end{table}

\begin{table}
{\small
\setlength{\tabcolsep}{1mm}
\begin{tabular}{|r|r|r|r|}

\hline
\hline
$T~(K)$&$\rho~(g/cm^3)$&$P\pm\delta P~(MPa)$&$U\pm\delta U~(KJ/mol)$\\
\hline
\hline

$240$&$1.05$&$22\pm 10$&$-47.09\pm 0.08$\\
$240$&$1.00$&$-106\pm 8$&$-46.81\pm 0.06$\\
$240$&$0.98$&$-163\pm 6$&$-47.48\pm 0.09$\\
$240$&$0.95$&$-209\pm 28$&$-46.52\pm 0.06$\\
$240$&$0.90^*$&$-305\pm 18$&$-46.87\pm 0.14$\\
$240$&$0.87$&$-342\pm 25$&$-45.82\pm 0.08$\\
$240$&$0.86^*$&$-329\pm 18$&$-46.23\pm 0.06$\\
$240$&$0.85$&$-268\pm 22$&$-45.78\pm 0.07$\\
$240$&$0.83$&$-236\pm 20$&$-46.42\pm 0.08$\\

\hline
$230$&$1.05$&$6\pm 4$&$-47.95\pm 0.03$\\
$230$&$1.00$&$-135\pm 5$&$-48.26\pm 0.03$\\
$230$&$0.98$&$-204\pm 27$&$-47.87\pm 0.09$\\
$230$&$0.95$&$-217\pm 8$&$-47.62\pm 0.09$\\
$230$&$0.90^*$&$-325\pm 24$&$-47.26\pm 0.09$\\
$230$&$0.87$&$-336\pm 20$&$-46.67\pm 0.07$\\
$230$&$0.86^*$&$-330\pm 10$&$-47.52\pm 0.10$\\
$230$&$0.85$&$-337\pm 17$&$-46.63\pm 0.08$\\
$230$&$0.83$&$-257\pm 23$&$-46.14\pm 0.09$\\

\hline
$220$&$1.05$&$-39\pm 33$&$-49.67\pm 0.08$\\
$220$&$1.00$&$-90\pm 15$&$-48.74\pm 0.08$\\
$220$&$0.98$&$-170\pm 25$&$-48.87\pm 0.09$\\
$220$&$0.95$&$-219\pm 5$&$-48.18\pm 0.05$\\
$220$&$0.90^*$&$-335\pm 24$&$-48.29\pm 0.08$\\
$220$&$0.87$&$-327\pm 26$&$-47.95\pm 0.05$\\
$220$&$0.86^*$&$-339\pm 14$&$-47.78\pm 0.08$\\
$220$&$0.85$&$-305\pm 11$&$-46.85\pm 0.07$\\
$220$&$0.83$&$-229\pm 26$&$-47.38\pm 0.05$\\

\hline
$210$&$1.05$&$9\pm 30$&$-50.09\pm 0.09$\\
$210$&$1.00$&$-107\pm 30$&$-49.71\pm 0.07$\\
$210$&$0.98$&$-176\pm 22$&$-49.53\pm 0.08$\\
$210$&$0.95$&$-226\pm 27$&$-49.150\pm 0.08$\\
$210$&$0.90^*$&$-393\pm 25$&$-48.88\pm 0.15$\\
$210$&$0.87$&$-352\pm 21$&$-48.39\pm 0.09$\\
$210$&$0.86^*$&$-253\pm 30$&$-49.10\pm 0.11$\\
$210$&$0.85$&$-263\pm 4$&$-47.56\pm 0.08$\\
$210$&$0.83$&$-337\pm 24$&$-47.86\pm 0.08$\\

\hline
\hline

\end{tabular}}
\caption{Temperature, density, pressure and internal energy of
the simulated state points.}
\label{table2}
\end{table}

\end{document}